\documentclass[%
 reprint,
superscriptaddress,
showpacs,preprintnumbers,
 amsmath,amssymb,
 aps,
prc,
]{revtex4-1}

\usepackage{graphicx}
\usepackage{dcolumn}
\usepackage{bm}
\usepackage{epstopdf}
\usepackage{multirow}
\usepackage{enumitem}
\usepackage{paralist}
\usepackage{booktabs}
\usepackage[version=3]{mhchem}
\usepackage{color}
\usepackage{placeins}
\usepackage{verbatim}

\begin{document}

\preprint{APS/123-QED}

\title{Spins and magnetic moments of $^{58,60,62,64}$Mn ground states and isomers}

\author{H. Heylen}
\email{hanne.heylen@fys.kuleuven.be}
\affiliation{KU Leuven, Instituut voor Kern- en Stralingsfysica, 3001 Leuven, Belgium}

\author{C. Babcock}
\email{carla.babcock@cern.ch}
\affiliation{Oliver Lodge Laboratory, Oxford Street, University of Liverpool, L69 7ZE, United Kingdom}
\affiliation{ISOLDE, Physics Department, CERN, CH-1211 Geneva 23, Switzerland}
\author{J. Billowes}
\affiliation{School of Physics and Astronomy, The University of Manchester, Manchester M13 9PL, United Kingdom}
\author{M. L. Bissell}
\affiliation{KU Leuven, Instituut voor Kern- en Stralingsfysica, 3001 Leuven, Belgium}
\author{K. Blaum}
\affiliation{Max-Plank-Institut f\"ur Kernphysik, D-69117 Heidelberg, Germany}
\author{P. Campbell}
\affiliation{School of Physics and Astronomy, The University of Manchester, Manchester M13 9PL, United Kingdom}
\author{B.~Cheal}
\affiliation{Oliver Lodge Laboratory, Oxford Street, University of Liverpool, L69 7ZE, United Kingdom}
\author{R. F. Garcia Ruiz}
\affiliation{KU Leuven, Instituut voor Kern- en Stralingsfysica, 3001 Leuven, Belgium}
\author{Ch.~Geppert}
\affiliation{Johannes Gutenberg-Universit\"at Mainz, Institut f\"ur Kernchemie, D-55128, Germany}
\affiliation{Institut f\"ur Kernphysik, TU Darmstadt, D-64289 Darmstadt, Germany}
\author{W. Gins}
\affiliation{KU Leuven, Instituut voor Kern- en Stralingsfysica, 3001 Leuven, Belgium}
\author{M. Kowalska}
\affiliation{ISOLDE, Physics Department, CERN, CH-1211 Geneva 23, Switzerland}
\author{K. Kreim}
\affiliation{Max-Plank-Institut f\"ur Kernphysik, D-69117 Heidelberg, Germany}
\author{S. M. Lenzi}
\affiliation{Dipartimento di Fisica e Astronomia dell'Universit\`a and INFN, Sezione di Padova, Padova, Italy}
\author{I.D.~Moore}
\affiliation{Department of Physics, University of Jyv\"askyl\"a, PB 35 (YFL) Jyv\"askyl\"a, Finland}
\affiliation{Helsinki Institute of Physics, FI-00014, University of Helsinki, Finland}
\author{R. Neugart}
\affiliation{Max-Plank-Institut f\"ur Kernphysik, D-69117 Heidelberg, Germany}
\affiliation{Johannes Gutenberg-Universit\"at Mainz, Institut f\"ur Kernchemie, D-55128, Germany}
\author{G. Neyens}
\affiliation{KU Leuven, Instituut voor Kern- en Stralingsfysica, 3001 Leuven, Belgium}
\author{W. N\"ortersh\"auser}
\affiliation{Institut f\"ur Kernphysik, TU Darmstadt, D-64289 Darmstadt, Germany}
\author{J. Papuga}
\affiliation{KU Leuven, Instituut voor Kern- en Stralingsfysica, 3001 Leuven, Belgium}
\author{D. T. Yordanov}
\email{Present address: Institut de Physique Nucl�aire d'Orsay, CNRS/IN2P3, Universit� Paris-Sud, F-91406 Orsay Cedex, France}
\affiliation{Max-Plank-Institut f\"ur Kernphysik, D-69117 Heidelberg, Germany}

\date{\today}

\begin{abstract}
The odd-odd $^{54,56,58,60,62,64}$Mn isotopes ($Z=25$) were studied using bunched-beam collinear laser spectroscopy at ISOLDE, CERN. From the measured hyperfine spectra the spins and magnetic moments of Mn isotopes up to $N=39$ were extracted. The previous tentative ground state spin assignments of $^{58,60,62,64}$Mn are now firmly determined to be $I=1$ along with an $I=4$ assignment for the isomeric states in $^{58,60,62}$Mn. The $I=1$ magnetic moments show a decreasing trend with increasing neutron number while the $I=4$ moments remain quite constant between $N=33$ and $N=37$. The results are compared to large-scale shell-model calculations using the GXPF1A and LNPS effective interactions. The excellent agreement of the ground state moments with the predictions from the LNPS calculations illustrates the need for an increasing amount of proton excitations across $Z=28$ and neutron excitations across $N=40$ in the ground state wave functions from $N=37$ onwards.

\end{abstract}

\maketitle

\section{\label{sec:Introduction}Introduction}
Ongoing developments in radioactive beam production provide the opportunity to study increasingly exotic nuclei. In recent years, a substantial effort has been devoted to understanding the nuclear structure of these exotic nuclei, in particular the often drastic variations in shell structure when moving from $\beta$-stability towards the drip lines. The neutron-rich $pf$-shell nuclei are of considerable interest because of the rapid structure evolution in the $Z<28$ and $N\approx 40$ region. 
Due to the stabilizing effect of the $Z=28$ shell closure, the ${}^{68}_{28}$Ni$_{40}$ ground state remains spherical whereas removal of only a few protons from this $Z=28$ shell results in a onset of deformation in ${}_{26}$Fe \cite{Hannawald1999a,Ljungvall2010, Rother2011} and ${}_{24}$Cr \cite{Baugher2012,Crawford2013}. This deformation is associated with the development of quadrupole correlations arising when neutrons are promoted across $N=40$. These neutron excitations are facilitated for open $\pi f_{7/2}$ nuclei due to the weakening of the $N=40$ subshell, suggested to result from the reduced proton-neutron interaction between protons in $\pi f_{7/2}$ and neutrons in $\nu f_{5/2}$ and $\nu g_{9/2}$ orbitals \cite{Otsuka2005}.
\\ Nuclear spins and moments are powerful probes to identify and understand such modifications of the nuclear structure due to their sensitivity to the composition of the wave function and deformation. These observables can be precisely extracted from high-resolution hyperfine spectra measured by collinear laser spectroscopy \cite{ChealFlanagan2010,Blaum2013}.
\\ At Jyv\"askyl\"a, this method was applied to measure hyperfine structures of Mn ($Z = 25$) isotopes near the $N = 28$ shell closure \cite{Charlwood2010}. At CERN-ISOLDE, this work has recently been extended to the odd-even Mn  isotopes up to $N=38$  \cite{Carla:prep}. The magnetic moments beyond $N=31$ were determined and the $I=5/2$ spins of $^{59,61,63}$Mn could be firmly established for the first time. From the systematic trend of the $g$-factors, $g=\mu/(I\mu_N)$, a structural change between $N=36$ and $N=38$ is inferred. The results were compared to large-scale shell-model calculations using the LNPS \cite{Lenzi2010} and GXPF1A \cite{Honma2004,Honma2005} effective interactions. The failure of the GXPF1A calculations to reproduce the neutron-rich $g$-factors, and the very good agreement of the LNPS predictions demonstrates the need for a model space including the $\nu g_{9/2}$ and $\nu d_{5/2}$ orbitals from $N=36$ onwards. Along with a gradual rise in the number of neutron excitations across $N=40$ with increasing $N$, an increase in proton excitations across $Z=28$ indicates that both proton and neutron excitations are important in the ground state wave function of Mn isotopes towards $N=40$. 
\\ In the current article, the ${\text{odd-odd}}$ $^{54,56,58,60,62,64}$Mn results obtained in the ISOLDE-experiment are presented. In the low-energy structure of $^{58,60,62,64}$Mn two long-lived states are known with tentatively assigned $I^\pi=(1^+)$ and $I^\pi=(4^+)$ spin-parity. Although firm spin-parity assignments are indispensable for the construction of reliable level schemes, prior to this work only indirect assignments based on $\beta$-decay measurements, shell-model calculations and regional systematics have been presented \cite{Gaudefroy2005,Liddick2006, Chiara2010,Liddick2011}. Furthermore, the nuclear configuration and degree of collectivity of these states is still unknown. Recently, Liddick \emph{et al.} suggested shape coexistence between a deformed $(1^+)$ state and spherical $(4^+)$ state in $^{64}$Mn \cite{Liddick2011}. These conclusions were based on the available information on neighboring nuclei and require verification by direct experimental evidence. For the lighter $^{58-62}$Mn isotopes only theoretical attempts to describe the nature of these states have been made. Sun \emph{et al.} described $^{58}$Mn and $^{60}$Mn in the projected shell-model framework using prolate deformed basis states \cite{Sun2012}. Alternatively, shell-model calculations of $^{57-62}$Mn were performed using an extended pairing-plus-quadrupole interaction with monopole corrections in the $fpg_{9/2}$ model space \cite{Jin2013}. In the latter study the significance of neutron excitations to the $\nu g_{9/2}$ intruder orbital has been pointed out. Although magnetic and quadrupole moments are key probes to investigate the nuclear structure of odd-odd Mn isotopes, no electromagnetic moments were published in these theoretical studies. 
\\ Here we present the direct measurement of spins and magnetic moments of the odd-odd Mn ground states from $N=29$ up to $N=39$ including the $^{58m,60m,62m}$Mn isomers.  We compare our results to large-scale shell-model calculations using the GXPF1A interaction with protons and neutrons restricted to the $pf$-orbits only, as well as to the LNPS interaction where neutron excitations into the positive-parity $\nu g_{9/2}$ and $\nu d_{5/2}$ orbitals are allowed.

\section{Experimental method and data analysis}
Radioactive manganese isotopes were produced at ISOLDE, CERN by bombarding a  thick UC$_x$ target with a 1.4 GeV proton beam. After the Mn atoms diffused out of the hot target, they were selectively ionized using the RILIS laser ion source \cite{RILIS}. Next, the 40 keV ion beam was mass separated in the high-resolution separator (HRS) and sent to the ISCOOL cooler-buncher \cite{ISCOOL} where, after each proton pulse, several 6 $\mu$s ion bunches were produced by a successive 43 ms accumulation and fast release. These ion bunches were guided to the dedicated collinear laser spectroscopy beam line COLLAPS, schematically shown in \cite{Papuga2014}, and overlapped with a co-propagating laser.
Following neutralization in a sodium vapor-filled charge exchange cell, the created atoms interact with the laser light and resonance fluorescence is observed in the subsequent interaction region. Laser light was obtained from a frequency-doubled narrowband dye laser to probe the $3d^5 4s^2\ {}^{6}S_{5/2} \rightarrow 3d^5 4s4p \  {}^{6}P_{3/2}$ ground state atomic transition (transition wavenumber: 35689.980~cm$^{-1}$) in manganese. The laser frequency was scanned across the hyperfine structure by applying a Doppler-tuning voltage to the ions prior to neutralization.
	\begin{figure}[t]
	\includegraphics[width=0.9\columnwidth]{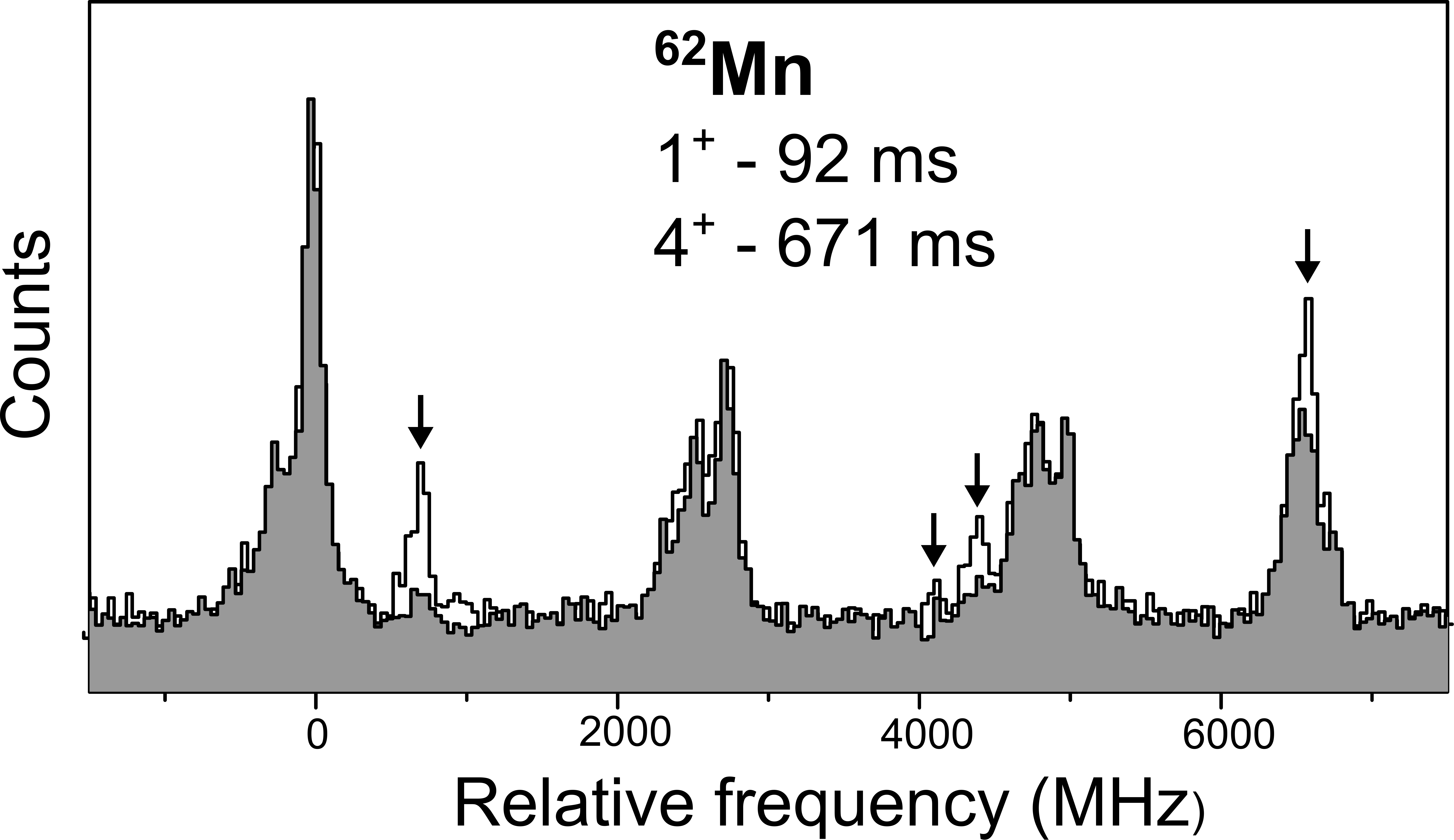}
	\caption{Hyperfine spectra of $^{62}$Mn taken without (shaded spectrum) and with (plain spectrum) proton gating. With proton gating photons are accepted only during one half-life of the shortest-lived isomer after proton impact. This clearly enhances the hyperfine structure of the short-lived state (indicated with arrows) with respect to the long-lived state. For visual comparison, the non-gated spectrum is scaled to match the background intensity. The frequency is shown relative to the $^{6}S_{5/2} \rightarrow {}^{6}P_{3/2}$ literature transition frequency of $^{55}$Mn \cite{Basar2003}. } 
	\label{Fig:Protontriggering-62Mn}
	\end{figure}
\\ Laser-induced fluorescence was detected by three PMTs in the light collection region. Since this fluorescence signal was only accepted when the atom bunch passed in front of the PMTs,  the continuous laser background was reduced by a factor of $10^4$ (ratio of bunch period and bunch length). Additional background suppression for short-lived states in $^{60,62,64}$Mn was achieved by considering only atom bunches which arrived within one half-life after proton impact. This proton gating eliminates background collection for bunches in which the short-lived state has already decayed. As a result, the features of the short-lived state in the spectrum are enhanced with respect to the long-lived state, facilitating their identification. This effect is illustrated for $^{62}$Mn in Fig.~\ref{Fig:Protontriggering-62Mn}, where spectra taken with and without proton-gating are compared. Taking into account the half-lives (Table \ref{Table:HFparam-moments}) and relative intensities of the two states in the hyperfine spectra, the isomeric ratio $N(1^+)/N(4^+)$ is estimated to be around 30\%.
	\begin{figure}[t]
	\includegraphics[width=\columnwidth]{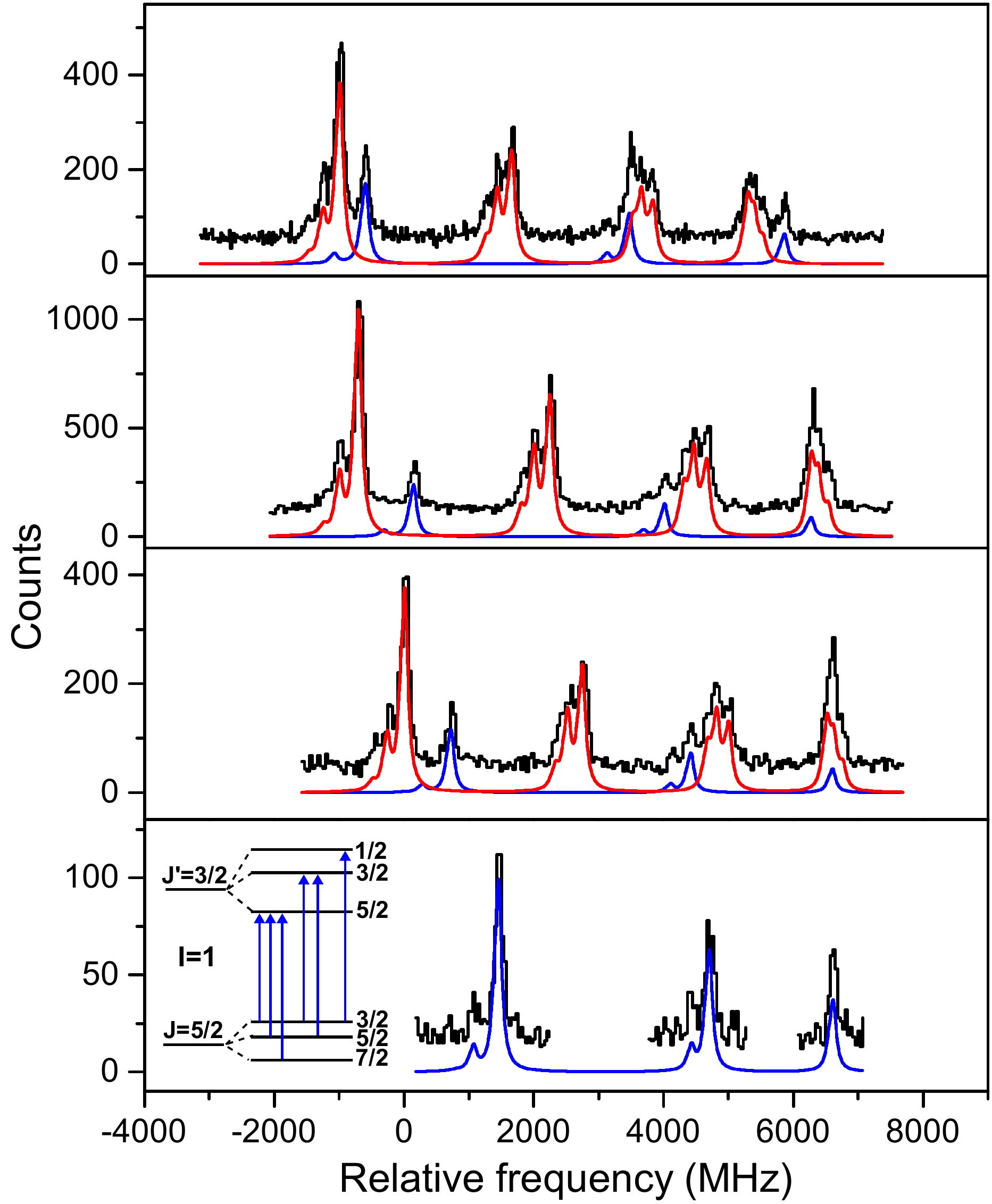}
	\caption{(Color online) Hyperfine spectra of the $^{58,60,62,64}$Mn $1^+$ states (blue) and $^{58,60,62}$Mn $4^+$ states (red) obtained by bunched-beam collinear laser spectroscopy on the atomic $^{6}S_{5/2} \rightarrow {}^{6}P_{3/2}$ line. For $^{60,62,64}$Mn proton gating was used. The inset shows a schematic depiction of the hyperfine structure of an $I=1$ state along with the allowed hyperfine transitions. The frequency is shown relative to the $^{6}S_{5/2} \rightarrow {}^{6}P_{3/2}$ literature centroid of $^{55}$Mn \cite{Basar2003}. } 
	\label{Fig:HFS-58-64}
	\end{figure}
\\ Typical experimental spectra for $^{58,60,62,64}$Mn can be found in Fig.~\ref{Fig:HFS-58-64}. The centroid frequency and hyperfine parameters $A$ and $B$ were extracted from the hyperfine structures via a $\chi^2$ minimization routine using the expression for hyperfine splitting energies (see e.g.\ \cite{ChealFlanagan2010}). $A(P_{3/2})$ and $B(P_{3/2})$ were left as free fit parameters while $B(S_{5/2})$ was set to 0 MHz considering our experimental uncertainty is not better than 0.2 MHz and the $^{55}$Mn literature value is $B(S_{5/2})=0.019 031(15)$ MHz \cite{Davis1971}. The hyperfine ratio was fixed to the value $A(P_{3/2})/A(S_{5/2}) =13.3$ established for $^{55}$Mn \cite{Carla:prep}. In order to correctly fit the partially unresolved peaks, the relative peak intensities were constrained to spin-dependent Racah values \cite{Magnante69}. It was verified that such constrained relative intensities do not significantly change the fit results as compared to fits with free intensities, which were only possible for isotopes without isomers. The peaks were fitted with Lorentzian line profiles with common widths for all peaks in the spectrum. To account for inelastic processes in the charge exchange cell \cite{Bendali1986,Kreim2014} which result in a small asymmetric line shape, a Lorentzian satellite peak at a fixed energy offset  (4.61(10) eV) was included. Due to the complex nature of the non-resonant ${\text{Mn}^+ - \text{Na}}$ charge exchange and the various atomic multiplets available, this energy offset was determined empirically. 

\section{Results}
In literature the spins of $^{58-64}$Mn were previously only tentatively assigned; $I=(1^+)$ and $I=(4^+)$ for the ground- and isomeric states were adopted in the most recent data compilations \cite{A58_2010,A60_2013,A62_2012,A64_2007} while earlier publications also reported $I=(0^+,2^+)$ and $I=(3^+)$ \cite{Ajzenberg1985,A58_1990,A60_2003,A62_2000}. As a result of our hyperfine spectra analysis, the spins can now be firmly established but no information on the relative energies of the states is obtained. In addition, from the hyperfine $A$ and $B$ parameters the magnetic and quadrupole moments are extracted. The sign and value of the deduced $g$-factor enable parity assignments.

	\begin{figure}[t]
	\includegraphics[width=0.95\columnwidth]{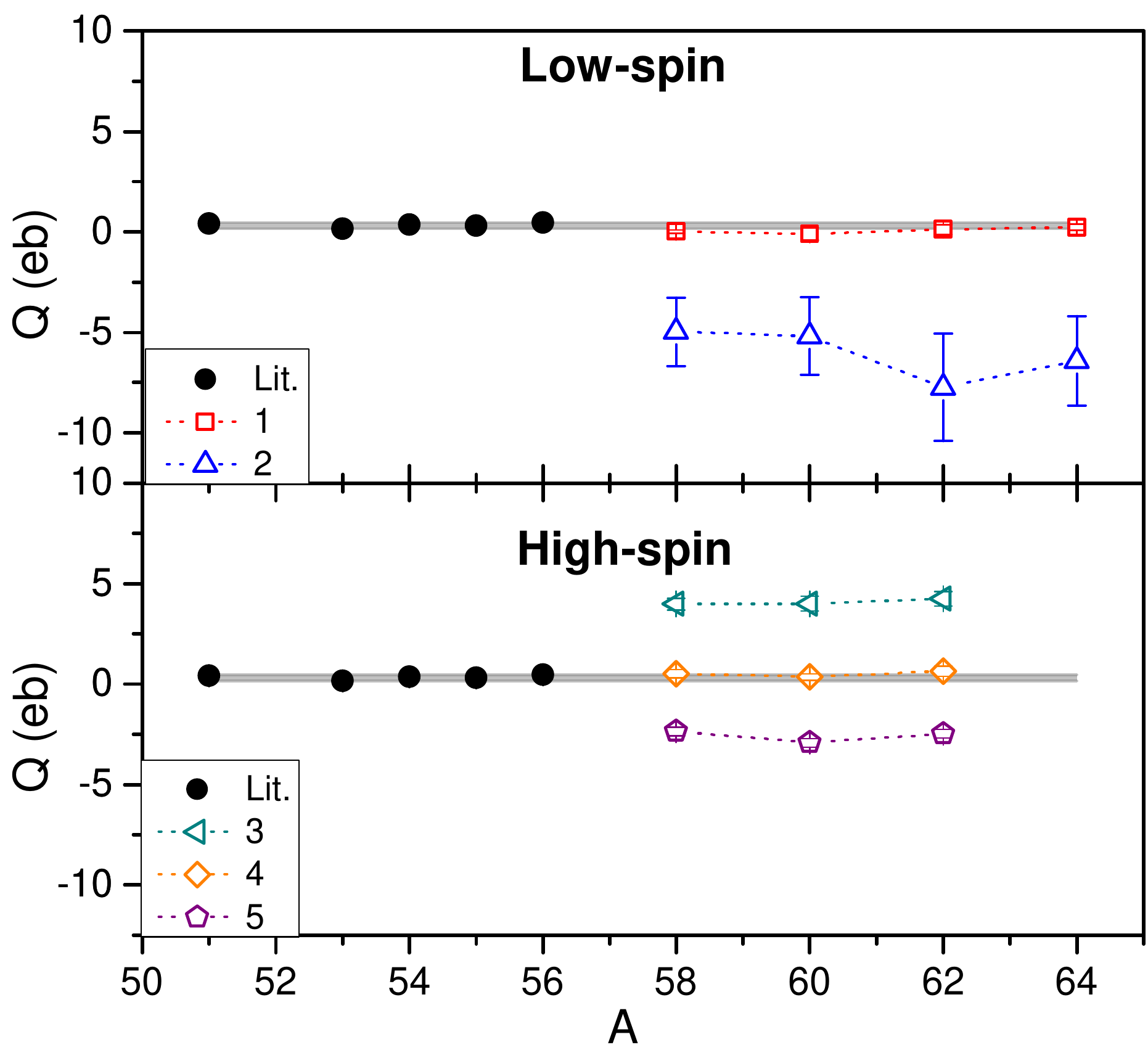}
	\caption{(Color online) Extracted quadrupole moments assuming different spins for the low-spin state and high-spin state in $^{58,60,62,64}$Mn. Note that the error bars are in most cases smaller than the symbols on this scale. These are compared with the known literature quadrupole moments. The shaded area represents the minimum and maximum literature value in the Mn chain.} 
	\label{Fig:Spins-quadrupolemoments}
	\end{figure}

\begin{table*}[t] 
\caption{Extracted spins,  hyperfine structure parameters $A(P_{3/2})$ and $B(P_{3/2})$ and corresponding magnetic moments and quadrupole moments for $^{54,56,58,60,62,64}$Mn ground states and isomers. The ratio of hyperfine parameters is constrained to $A(P_{3/2})/A(S_{5/2})=13.3$ found in the odd Mn isotopes \cite{Carla:prep}. No hyperfine anomaly correction is taken into account for the calculation of the magnetic moments.}
\begin{ruledtabular}
\begin{tabular}{c c r r r r r r r }
\\ [-2.3ex]
$A$ & $I^{\pi}$ & $T_{1/2}$ & $A(P_{3/2})$ (MHz) & $B(P_{3/2})$ (MHz) & $\mu_\text{exp}$ $(\mu_\text{N})$ & $\mu_\text{lit}$ $(\mu_\text{N})$ & $Q_{s,\text{exp}}$ ($e$b) & $Q_{s,\text{lit}}$ ($e$b) \\
 [0.45ex]
\colrule \\ [-1.85ex]
54 & 3$^+$ & 312 d & -763.2\,(7) & +15\,(6) & +3.299\,(3)&   +3.2966\,(2)\footnotemark[1]& +0.37\,(14) &  +0.33\,(3)\footnotemark[1]  \\
 &  &  & && &   +3.306\,(1)\footnotemark[2]&  &  +0.37\,(3)\footnotemark[2]  \\
56 & 3$^+$ & 2.59 h & -750.2\,(4) & +10\,(3) & +3.242\,(2) & +3.24121\,(13)\footnotemark[3]  &+0.24\,(8) & +0.47(15)\footnotemark[2] \\
58 & 1$^+$ & 3.0 s & -1821.5\,(16) & +1\,(3) & +2.624\,(2) && +0.03\,(8) & \\
 & 4$^+$ & 65.4 s& -531.7\,(3) & +20\,(4) & +3.064\,(2) && +0.47\,(9)& \\
60 & 1$^+$ & 0.28 s & -1728\,(5) & +4\,(9) & +2.489\,(7) & &+0.1\,(2) & \\
 & 4$^+$ & 1.77 s& -590.3\,(3) & +14\,(4) & +3.402\,(2) & &+0.33\,(10) & \\
62 & 1$^+$ & 92 ms &-1656\,(5) & +4\,(9) & +2.385\,(7) & &+0.1\,(2) & \\
 & 4$^+$ & 671 ms & -550.4\,(4) & +25\,(5) & +3.172\,(3) & & +0.59\,(13)& \\
64 & 1$^+$ & 90 ms & -1448\,(3) & +9\,(5) & +2.086\,(3) & & +0.21\,(11)& \\
 & (4$^+$) & 0.50 ms & - & - & - & & - &  \\
 [1ex]
\emph{55} & \emph{5/2} & \emph{stable} & \emph{-963.1(3)} & \emph{+13.4(8)} & \multicolumn{1}{c}{\emph{Ref.}} & \emph{+3.4687179(9)\footnotemark[4]} &\multicolumn{1}{c}{\emph{Ref.}} &  \emph{0.33(1)\footnotemark[5]}\\
\end{tabular}
\end{ruledtabular}
\footnotetext[1]{Nuclear magnetic resonance on oriented nuclei data \cite{Niesen1970} recalculated with current $^{55}$Mn reference values}
\footnotetext[2]{Collinear laser spectroscopy experiment on the $S_2 \rightarrow P_3$ ionic transition \cite{Charlwood2010}} 
\footnotetext[3]{Atomic beam magnetic resonance data \cite{Childs1961} recalculated with current $^{55}$Mn reference values.}
\footnotetext[4]{Nuclear magnetic resonance data \cite{Lutz1974} corrected for diamagnetic shielding.}
\footnotetext[5]{Atomic beam magnetic resonance experiment \cite{Dembczynski1979}.}
\label{Table:HFparam-moments}
\end{table*}	
\subsection{Low-spin state: $I=1$} 
Despite the predominant high-spin state, multiple peaks corresponding to the low-spin state can be clearly identified in the hyperfine spectra of Fig. \ref{Fig:HFS-58-64}. These multiple peaks are in contradiction with the absence of hyperfine splitting in the case of $I=0$ which would yield only one peak. An $I=0$ spin assignment can therefore be ruled out unambiguously. To definitely decide between $I=1$ and $I=2$, both spin assignments were considered in the initial analysis. From the hyperfine transition selection rules, three hyperfine multiplets are expected in the case of $I=1$ while four multiplets should appear for $I=2$. Although no evidence for a fourth multiplet is seen on the wide frequency scans, the possibility cannot be disregarded considering the rather low statistics. Therefore, for $I=2$, the spectra are fitted assuming one multiplet is unaccounted for on the high frequency side of our measuring range. An unaccounted multiplet on the left can be ruled out, as this would severely violate the smooth increasing trend in centroid frequency seen in the Mn chain.
\\ As shown in Fig. \ref{Fig:Spins-quadrupolemoments}, the extracted quadrupole moments assuming an $I=2$ in the fitting procedure are an order of magnitude larger than the known values in the Mn chain and such large quadrupole moments are found only occasionally in much heavier mass regions (details on $Q$-determination are given in section \ref{sec:HFparam}). These extraordinarily large quadrupole moments combined with a worse fit quality (reflected in a higher $\chi^2_\text{red}$) exclude a possible $I=2$ spin assignment.
\\ Based on the presented evidence, we firmly assign $I=1$ for all low-spin states in $^{58,60,62,64}$Mn.

\subsection{High-spin state: $I=4$}
The different $I=3,4$ and 5 spin possibilities all yield the same number of multiplets in the hyperfine spectrum, hence it is impossible to discriminate between the spins based on this criterion. From fitting the high-spin structures in $^{58,60,62}$Mn with the alternative spins, quadrupole moments can be extracted as shown in Fig. \ref{Fig:Spins-quadrupolemoments}. Analogous to the low-spin case, the unrealistically large moments in the case of $I=3$ and 5 exclude these spins, establishing $I=4$ in $^{58,60,62}$Mn. The high-spin state in $^{64}$Mn could not be measured because of its short 0.50 ms half-life.
This $I=4$ spin in combination with the $I=1$ low-spin assignment agrees with the earlier suggested $M3$ multipolarity of the isomeric transitions in $^{58,60}$Mn, measured via electron conversion spectroscopy \cite{Schmidtott1993}.

\subsection{Hyperfine parameters}
\label{sec:HFparam}
The experimental hyperfine parameters obtained with the above established spins, along with the corresponding nuclear moments are presented in Table \ref{Table:HFparam-moments}. The nuclear moments were extracted by comparison with the precise $^{55}$Mn reference moments \cite{Lutz1974,Dembczynski1979} and hyperfine parameters \cite{Carla:prep} using \begin{align*}
\mu =  \mu_\text{ref} \frac{IA}{I_\text{ref}A_\text{ref}} \quad\text{and}\quad Q_{s} = Q_{s,\text{ref}}\frac{B}{B_\text{ref}}.
\end{align*} 
The hyperfine anomaly is assumed to be negligible.
For $^{54,56}$Mn there is a satisfactory agreement between our results and the previously known literature values \cite{Niesen1970,Charlwood2010, Childs1961}.
\\ Since the quadrupole splitting of both the lower and the upper atomic state is small, the quadrupole moment sensitivity is low, resulting in large relative uncertainties. Due to this low precision, a detailed discussion on the quadrupole moments is not presented. It should be noted however that the consistently smaller quadrupole moment of the $I=1$ state compared to the $I=4$ state does not imply that the latter is intrinsically more deformed. Under the assumption that the nuclear deformation is axially symmetric, the intrinsic quadrupole moment $Q_0$ can be related to the measured moment $Q_s$ using 
\begin{equation*}
Q_s = \frac{3K^2 - I(I+1)}{(I+1)(2I+3)}Q_0,
\end{equation*}
where $K$ is the projection of the total spin $I$ on the symmetry axis \cite{Neyens2003}. In the case that the  spin is along the symmetry axis ($K = I$), an identical intrinsic quadrupole moment results in a spectroscopic quadrupole moment which is 5 times smaller for $I=1$ than for $I=4$. With the current experimental precision, the smaller quadrupole moment for $I=1$ compared to $I=4$ can therefore be fully explained by this spin-factor. Furthermore, the presence of a large ground state deformation in $^{64}$Mn, as suggested by Liddick \emph{et al.} \cite{Liddick2011}, can not be confirmed nor refuted. A more precise measurement of the $1^{+}$ ground state quadrupole moments is needed to draw such a conclusion.

\subsection{Parity}
At present, a positive parity for the $I=1$ state in $^{58,60,62,64}$Mn  is proposed based on $\log ft$ values found in $\beta$-decay from the $0^+$ state in Cr \cite{Gaudefroy2005,Liddick2011} or $\beta$-decay to the $0^+$ state in Fe \cite{Bosch1988,Liddick2006}. Together with the observed $M3$ transitions, this suggest also a positive parity of the $I = 4$ state in $^{58,60}$Mn. However for $^{62,64}$Mn, the multipolarity of the isomeric transition is not measured experimentally and little $\beta$-decay data is available. Although the parity cannot be extracted from the hyperfine spectra directly, the sign and magnitude of the extracted $g$-factor $g =\mu /(I\mu_N$), can help its determination.
	 \begin{figure}[b]	
	\includegraphics[width=0.5\columnwidth]{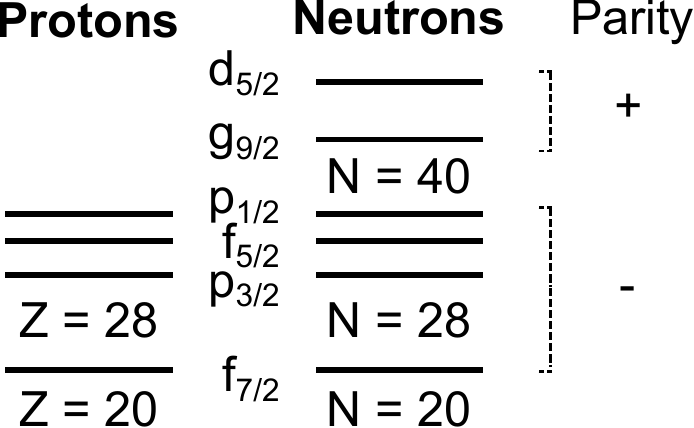}
	\caption{The single-particle orbitals and their parity relevant in the discussion of neutron-rich Mn isotopes. In a simple shell-model picture the 5 valence protons 		reside in $\pi f_{7/2}$ and the valence neutrons occupy the $pf$-orbitals.}
	\label{Fig:Modelspace}
	\end{figure}
\begin{table}[t]  
\caption{Summary of $g$-factors calculated with the
additivity rule using empirical $g$-factors. The configurations not yielding the spin of interest are indicated by -.}
\begin{ruledtabular}
\begin{tabular}{l l  r r }
\\ [-2.3ex]
Protons & Neutrons & $ I = 1$ & $I=4$ \\
 [0.45ex]
\colrule \\ [-1.85ex]
\multicolumn{4}{c}{Positive parity} \\
\colrule \\ [-1.85ex]
$\pi \left(f_{7/2}^{-3}\right)_{5/2}$ & $\nu p_{3/2}$ & +2.79 & +0.68 \\
  & $\nu f_{5/2}$ & +0.83 & +0.83 \\
  & $\nu p_{1/2}$ &\multicolumn{1}{c}{-}  & \multicolumn{1}{c}{-}\\
$\pi p_{3/2}$ & $\nu p_{3/2}$ & +0.70 & +0.70 \\
  & $\nu f_{5/2}$ & -0.93 & +0.89 \\
  & $\nu p_{1/2}$ & +2.07 & \multicolumn{1}{c}{-}  \\
$\pi f_{5/2}$ & $\nu p_{3/2}$ & +1.08 & +0.06 \\
  & $\nu f_{5/2}$ & +0.34 & +0.34 \\
  & $\nu p_{1/2}$ &\multicolumn{1}{c}{-}  &\multicolumn{1}{c}{-}  \\
 [0.45ex]
\colrule \\ [-1.85ex]
\multicolumn{4}{c}{Negative parity} \\
\colrule \\ [-1.85ex]
$\pi \left(f_{7/2}^{-3}\right)_{5/2}$ & $\nu g_{9/2}$ &\multicolumn{1}{c}{-}  & -0.09 \\
  & $\nu d_{5/2}$ & +0.04 & +0.04 \\
$\pi p_{3/2}$ & $\nu g_{9/2}$ & \multicolumn{1}{c}{-}  & -0.31 \\
 & $\nu d_{5/2}$ & -3.70 & -0.10 \\
$\pi f_{5/2}$ & $\nu g_{9/2}$ & \multicolumn{1}{c}{-} & -0.16 \\
 & $\nu d_{5/2}$ & -0.04 & -0.04 \\
\end{tabular}
\end{ruledtabular}
\label{Table:Parity}
\end{table}
Considering the relevant shell-model orbitals in Fig.~\ref{Fig:Modelspace}, the natural parity for an odd-odd nucleus is positive, a result of the coupling between protons and neutrons occupying the $pf$ orbitals. On the other hand, a negative parity state can only come from the coupling to an unpaired neutron in the $\nu g_{9/2}$ or $\nu d_{5/2}$ orbitals. For pure configurations arising from a weak coupling between the unpaired protons and neutron, the $g$-factor can be estimated via the $g$-factor additivity rule \cite{Heyde_Shellmodel}. For neutron-rich Mn this assumption is too simple and configuration mixing needs to be taken into account. Nevertheless, the estimated $g$-factor can be an indication of the sign and magnitude of the $g$-factor to be expected. For the calculations, shown in Table \ref{Table:Parity}, empirical $g$-factors are taken from $^{55}$Mn ($\pi f_{7/2}^{-3}$), $^{69}$Cu ($\pi p_{3/2}$), $^{75}$Cu ($\pi f_{5/2}$) for protons and $^{57}$Ni ($\nu p_{3/2}$), $^{65}$Ni ($\nu f_{5/2}$),  $^{67}$Ni ($\nu p_{1/2}$), $^{69}$Zn ($\nu g_{9/2}$), $^{91}$Zr ($\nu d_{5/2}$) for neutrons. As found in the odd Mn isotopes \cite{Carla:prep}, the leading proton configuration along the isotopic chain are three proton holes in $\pi f_{7/2}$ coupled to $I^{\pi}=5/2^-$, while for isotopes near $N=40$ also a contribution of the $\pi p_{3/2}$ and $\pi f_{5/2}$ orbitals cannot be disregarded.

\noindent The calculated $g$-factor for all  negative parity states is negative with the exception of the $\pi f_{7/2}^{-3} \otimes \nu d_{5/2}$  configuration where it is marginally positive with a value of +0.04. The experimental $g$-factors for both the $I=1$ ($+2.087< g< +2.624$) and $I=4$ ($+0.766< g< +0.850$) states are however positive and much larger than the calculated negative parity values. This strongly supports a positive parity assignment for the $I=1$ and $I=4$ states in $^{58,60,62,64}$Mn.

\section{\label{sec:Discussion}Discussion}
	\begin{figure*}[t]
	\includegraphics[width=2\columnwidth]{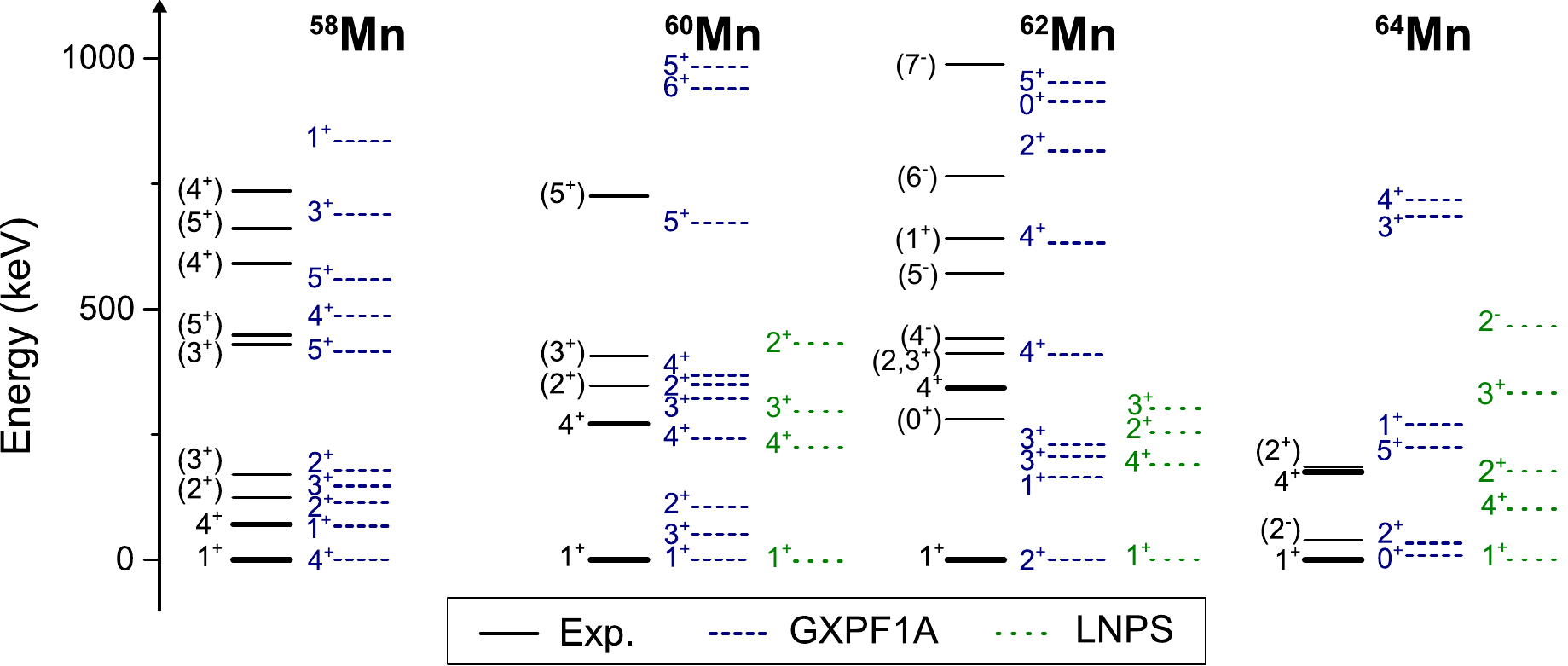}
	\caption{(Color online) $^{58,60,62,64}$Mn energy levels below 1 MeV compared to GXPF1A and LNPS calculations. The long-lived $1^+$ and $4^+$ states are highlighted with a bold line. Experimental levels taken from the most recent Nuclear Data Sheets \cite{A58_2010,A60_2013,A62_2012,A64_2007} and \cite{Liddick2011,Gaffney2015}.} 
	\label{Fig:Levelschemes}
	\end{figure*}
Shell-model calculations using two different effective interactions and model spaces have been performed to obtain energy levels, spins and magnetic moments. The single-particle orbitals relevant for the discussion below are schematically shown in Fig. \ref{Fig:Modelspace}.
	
\noindent The GXPF1A \cite{Honma2004,Honma2005} and LNPS \cite{Lenzi2010} effective interactions have been used with the ANTOINE shell-model code \cite{ANTOINE}. The GXPF1A model space comprises the full $pf$-shell for protons and neutrons with $^{40}$Ca as a core. Since the neutron $\nu g_{9/2}$ and $\nu d_{5/2}$ orbitals are not included, no excitations across $N=40$ are allowed. The model space in the LNPS calculations consists of the full $pf$-shell for protons and the $p_{3/2},f_{5/2},p_{1/2},g_{9/2}$ and $d_{5/2}$ orbitals for neutrons. This model space is unsuited for isotopes below $N=36$ due to the absence of cross-shell excitations across $N=28$ using a $^{48}$Ca core.  Due to the large dimensions of the model space, the Mn calculations require truncations. In the GXPF1A calculations up to  2 protons and 6 neutrons have been allowed to excite from $f_{7/2}$ to the higher orbitals while in the LNPS calculations a maximum of $11p-11h$ excitations across $Z=28$ and $N=40$ are allowed.

\noindent The currently available spectroscopic information below 1 MeV is compared to the shell-model predictions in Fig. \ref{Fig:Levelschemes}. Due to the lengthy computation time, only a few relevant low-spin states in $^{60,62,64}$Mn were calculated with LNPS. 
\\ The GXPF1A calculations reproduce the high level density in $^{58}$Mn, but  the agreement for the heavier isotopes is poor. The inversion of the $4^+$ and $1^+$ state at $^{58}$Mn is not significant considering the 119 keV rms energy deviation \cite{Steppenbeck2010}. Note that the spin-parity assignments of excited states in $^{58,60}$Mn are mostly based on the good correspondence between calculated and experimental level energies \cite{Steppenbeck2010}. Taking into account a typical $100 - 200$~ keV rms energy deviation, LNPS correctly reproduces the positive parity level sequence in $^{60,62,64}$Mn, while the energy of the isomeric $4^+$ state is predicted slightly lower than the experimental value. The energy of the low-lying $(2^-)$ negative parity state in $^{64}$Mn, arising due to the coupling with a neutron in the positive parity $gd$-orbits, is calculated almost 500 keV too high.
	\begin{figure}[t]
	\includegraphics[width=0.9\columnwidth]{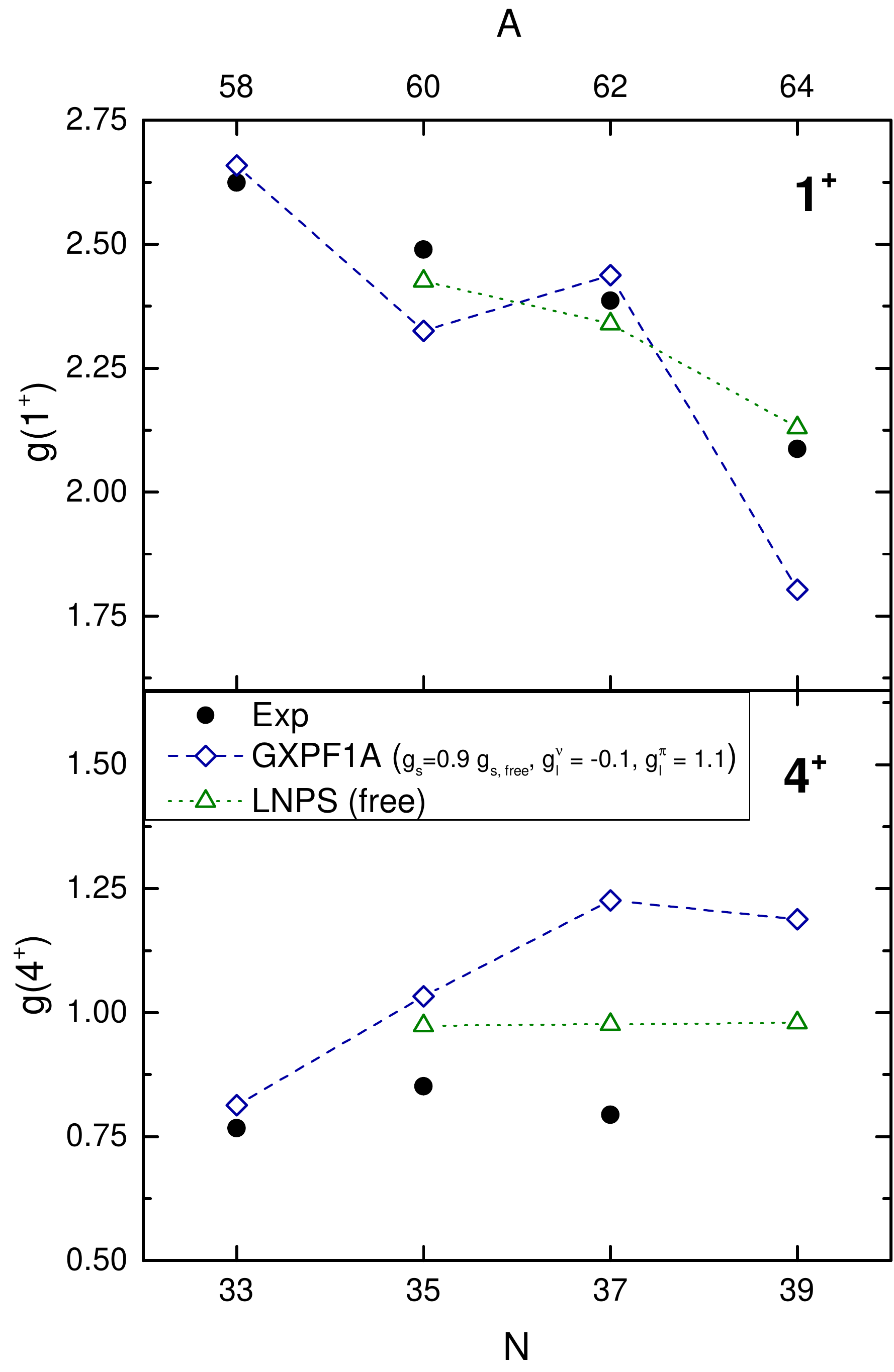}
	\caption{(Color online) $^{58,60,62,64}$Mn $g$-factors of the $1^+$ (top) and $4^+$ states (bottom). Comparison of experimental values with shell-model calculations performed with the GXPF1A (dashed line) and LNPS (dotted line) effective interactions using free $g$-factors.} 
	\label{Fig:gfactors}
	\end{figure}

\noindent In Fig. \ref{Fig:gfactors}, the experimental and predicted $g$-factors are presented. The calculations have been performed using free-nucleon spin and orbital $g$-factors for the LNPS interaction and mildly-quenched values for GXPF1A ($g_{s, \text{eff}}=0.9g_{s,\text{free}}$, $g^{\pi}_{l,\text{eff}}=1.1, g^{\nu}_{l,\text{eff}}=-0.1$) as in \cite{Honma2004}. The $1^+$ $g$-factors are excellently described by LNPS for $^{60,62,64}$Mn, whereas GXPF1A is unable to reproduce the downward slope, although there is a good agreement for $^{58}$Mn and $^{62}$Mn. For the $4^+$ state, GXPF1A correctly calculates the value for $^{58}$Mn, but fails to predict the nearly constant trend towards the more neutron-rich isotopes. On the other hand, the LNPS calculations do reproduce the rather constant $g$-factor although the actual value is about 20\% higher than the experimental value. 
	\begin{figure}[t]
	\includegraphics[width=0.9\columnwidth]{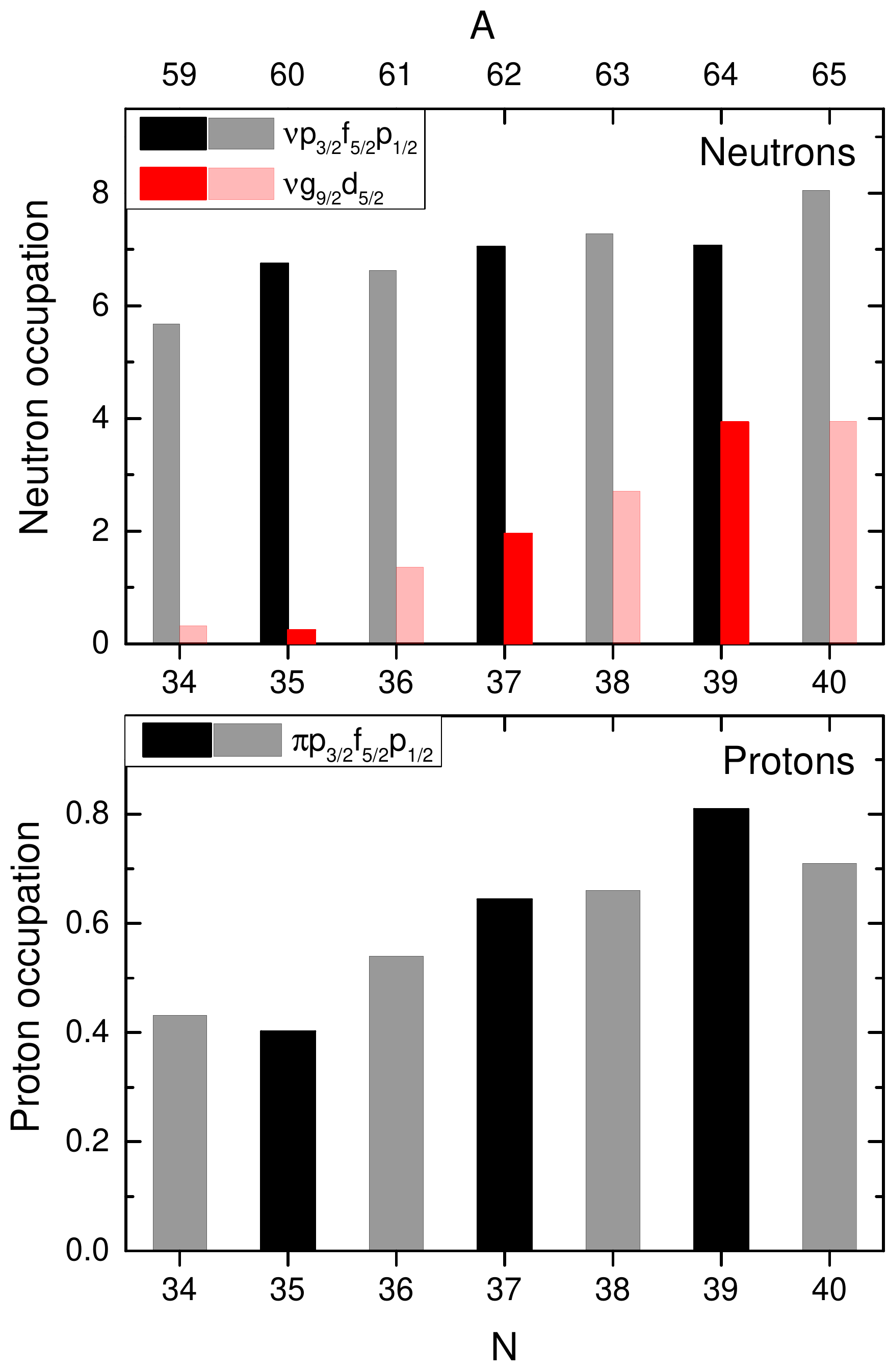}
	\caption{(Color online) Neutron (top) and proton (bottom) occupation numbers of the $1^+$ state calculated with the LNPS effective interaction. There is a large similarity between the odd-even Mn (shaded) and odd-odd Mn (solid) occupations.} 
	\label{Fig:Occupationnumbers-oddvsodd-odd}
	\end{figure}
\\ Using the empirically calculated $g$-factors from Table \ref{Table:Parity}, an attempt can be made to at least qualitatively explain the experimentally observed trends. The $g(1^+)$ at $^{58}$Mn could correspond to a configuration dominated by $\pi f_{7/2}^{-3} \otimes \nu p_{3/2}$ while the decreasing $g$-factor with increasing neutron number is then evidence for increased configuration mixing with other neutron orbitals in the $pf$-shell or admixtures with proton intruder states. Since all of these configurations would yield a smaller $g$-factor (see Table \ref{Table:Parity}), it is difficult to identify which of them are the main contributors. Note that although the $g$-factors clearly suggest a changing configuration, the spin remains $1^+$ for all four isotopes between $N=33$ and $N=39$.  The $g$-factors of the $4^+$ states are not very sensitive to the wave function configuration, as can be seen from the very similar $g$-factors for all configurations in Table \ref{Table:Parity}. This means that the nearly constant $g$-factor measured experimentally does not imply that also the nuclear configuration remains unchanged. 
\\ Although these calculations are valuable to build an intuitive interpretation of the $g$-factors, they are too simple to provide a quantitative description, especially considering the very fragmented wave functions predicted by the shell-model calculations. Therefore it is more instructive to look at calculated average proton and neutron occupation numbers in the different orbits. In Fig. \ref{Fig:Occupationnumbers-oddvsodd-odd}, these occupation numbers predicted by the LNPS interaction are shown for the $1^+$ state. Because of the sensitivity of $g$-factors to the wave function composition, the excellent correspondence between the experimental $g(1^+)$ of $^{60-64}$Mn suggests that the leading configurations of these $1^+$ states are calculated correctly. In the top panel, the increase in neutron occupation of the $\nu g_{9/2}$ and $\nu d_{5/2}$ orbitals (in red) towards $N=40$ can be seen while the occupation of the $\nu p_{3/2}\,f_{5/2}\,p_{1/2}$ orbitals (in black) remains almost constant at $7-8$, well below the maximum of 12 neutrons. This shows that from $N=37$ onwards the wave function is dominated by $2p-2h$ and $4p-4h$ neutron excitations across $N=40$, analogous to the odd Mn case \cite{Carla:prep}. 
The importance of neutron excitations to the $\nu g_{9/2}$ and $\nu d_{5/2}$ orbitals is further visible in the energy spectrum of $^{64}$Mn where a negative parity $2^-$  state appears between the natural parity $1^+$ and $4^+$ states. As a consequence of this intermediate $2^-$ state, an $M2-E1$ isomeric cascade is favored over the much slower $\beta$-decay, resulting in a half-life significantly shorter than those of the other studied $4^+$ states \cite{Liddick2011}. 
\\ In addition to neutron excitations, proton excitations across $Z=28$ become increasingly important towards $N=40$ as illustrated by the rising $\pi p_{3/2}\,f_{5/2}\,p_{1/2}$ occupation numbers in the bottom panel of Fig. \ref{Fig:Occupationnumbers-oddvsodd-odd}. Such a parallel rise of proton and neutron excitations is also seen in the odd Mn chain \cite{Carla:prep} and is understood as arising from the complex interplay between the protons in the $pf$ shell and neutrons in the $pf$ and $gd$ shells. More neutrons in the $\nu g_{9/2}$ and $\nu d_{5/2}$ orbitals result in a less bound proton $\pi f_{7/2}$ and more bound proton $\pi f_{5/2}$ orbital due to the proton-neutron tensor force \cite{Otsuka2005}. This effectively reduces the $Z=28$ shell gap, facilitating particle-hole excitations which lead to stronger deformation and, in turn, to more neutrons in $\nu g_{9/2}$ and $\nu d_{5/2}$. In contrast to the odd Mn case however, the trend in $g$-factors is smooth and no discontinuity related to a sudden structural change is seen \cite{Carla:prep}.
\\ For the $4^+$ states, LNPS predicts a similar increase in proton and neutron cross shell excitations as for the $1^+$ states. However, taking into account the 20\% overestimation of the calculated $g$-factor, other observables are desired to support this interpretation.

\section{Conclusion}
The hyperfine spectra of odd-odd Mn from $A=54$ up to $A=64$ were measured using collinear laser spectroscopy yielding  direct information on spins and magnetic moments. The previously suggested $I=1$ and $I=4$ spins for the long-lived states in $^{58,60,62,64}$Mn were firmly determined, except for the high-spin state in $^{64}$Mn which could not be observed due to its short half-life.
\\ The experimental results are compared with shell-model calculations using two different effective interactions in a model space restricted to the $pf$ shell (GXPF1A), and in a model space including the $\nu g_{9/2}$ and $\nu d_{5/2}$ orbitals (allowing excitations across $N=40$) but excluding the $\nu f_{7/2}$ (LNPS). A comparison of energy levels, spins and $g$-factors shows that a model space without neutron excitations across $N=40$ is too limited to accurately predict observables of the neutron-rich Mn isotopes. On the other hand, calculations with the LNPS interaction, including such excitations, give an excellent description of the $1^+$ $g$-factors. In addition, the flat trend of the $g$-factors as a function of the neutron number of the $4^+$ states is correctly reproduced although the actual value is overestimated by about 20\%.
From the average occupation numbers in LNPS, the importance of both neutron excitations across $N=40$ as well as proton excitations across $Z=28$ is established.
\\ The determination of spins and magnetic moments is an important first step in understanding the changing nuclear structure in the neutron-rich Mn isotopes near $N=40$ but for direct information on possible shape transitions, a more precise knowledge of the quadrupole moments is required, while also the mean square charge radii will provide complementary information.

\section*{Acknowledgements}

We would like to thank the ISOLDE technical group for their support and assistance. This work was supported by the Belgian Research
Initiative on Exotic Nuclei (IAP-project P7/12), the FWO-Vlaanderen, GOA grant 10/010 and 10/015 from KU Leuven, NSF Grant No. PHY-1068217, BMBF (05 P12 RDCIC), the Max-Planck Society, the Science and Technology Facilities Council,
and EU FP7 via ENSAR (No. 262010).

\bibliography{Mn}

\end{document}